\documentstyle[aps,prl,twocolumn,epsf]{revtex}
\begin{document}
\input{psfig}
\draft
\twocolumn[\hsize\textwidth\columnwidth\hsize\csname 
@twocolumnfalse\endcsname 

\title{Soft Phonon Anomalies in the Relaxor Ferroelectric
  Pb(Zn$_{1/3}$Nb$_{2/3}$)$_{0.92}$Ti$_{0.08}$O$_3$}

\author{ P.\ M.\ Gehring$^{(1)}$, S.\ -E.\ Park$^{(2)}$, G.\ Shirane$^{(3)}$ }

\address{ $^{(1)}$NIST Center for Neutron Research, National Institute
of Standards and Technology, Gaithersburg, Maryland 20899 }

\address{ $^{(2)}$Materials Research Laboratory, The Pennsylvania
State University, University Park, Pennsylvania 16802 }

\address{ $^{(3)}$Physics Department, Brookhaven National Laboratory,
Upton, New York 11973 }

\date{\today}
\maketitle

\begin{abstract}
  Neutron inelastic scattering measurements of the polar TO phonon
  mode dispersion in the cubic relaxor
  Pb(Zn$_{1/3}$Nb$_{2/3}$)$_{0.92}$Ti$_{0.08}$O$_3$ at 500~K reveal
  anomalous behavior in which the optic branch appears to drop
  precipitously into the acoustic branch at a finite value of the
  momentum transfer $q = 0.2$~\AA$^{-1}$ measured from the zone
  center.  We speculate this behavior is the result of nanometer-sized
  polar regions in the crystal.
\end{abstract}


]

The discovery by Kuwata {\it et al.} in 1982 that it was possible to
produce single crystals of the relaxor-ferroelectric material
Pb(Zn$_{1/3}$Nb$_{2/3}$)$_{1-x}$Ti$_x$O$_3$ represented an important
achievement in the field of ferroelectrics \cite{Kuwata}.  Because the
parent compounds Pb(Zn$_{1/3}$Nb$_{2/3}$)O$_3$ (PZN) and PbTiO$_3$
(PT) form a solid solution, it was possible to tune the stoichiometry
of the material to lie near the morphotropic phase boundary (MPB) that
separates the rhombohedral and tetragonal regions of the phase diagram
\cite{Kuwata,Park1,Park2}.  Such MPB compositions in
Pb(Zr$_{1-x}$Ti$_x$)O$_3$ (PZT), the material of choice for the
fabrication of high-performance electromechanical actuators, exhibit
exceptional piezoelectric properties, and have generated much
scientific study \cite{Noheda}.  However, in contrast to PZN-$x$PT,
all attempts to date to grow large single crystals of PZT near the MPB
have failed, and this has impeded progress in fully characterizing the
PZT system.

The dielectric and piezoelectric properties of single crystals of both
PZN-$x$PT and PMN-$x$PT (M = Mg) have since been examined by Park {\it
  et al.} who measured the strain as a function of applied electric
field \cite{Park1,Park2}.  These materials were found to exhibit
remarkably large piezoelectric coefficients $d_{33} > 2500$~pC/N and
strain levels $S \sim$ 1.7\% for rhombohedral crystals oriented along
the pseudo-cubic [001] direction.  This level of strain represents an
order of magnitude increase over that presently achievable by
conventional piezoelectric and electrostrictive ceramics including
PZT.  That these ultrahigh strain levels can be achieved with nearly
no dielectric loss ($< 1$\%) due to hysteresis suggests both PMN-$x$PT
and PZN-$x$PT hold promise in establishing the next generation of
solid state transducers \cite{Service}.  A very recent theoretical
advance in our understanding of these materials occurred when it was
shown using first principles calculations that the {\it intrinsic}
piezoelectric coefficient $e_{33}$ of MPB PMN-40\%PT was dramatically
enhanced relative to that for PZT by a factor of 2.7 \cite{Bellaiche}.
Motivated by these experimental and theoretical results, we have
studied the dynamics of the soft polar optic phonon mode in a high
quality single crystal of PZN-8\%PT, for which the measured value of
$d_{33}$ is a maximum, using neutron inelastic scattering methods.

In prototypical ferroelectric systems such as PbTiO$_3$ it is well
known that the condensation or softening of a zone-center transverse
optic (TO) phonon is responsible for the transformation from a cubic
paraelectric phase to a tetragonal ferroelectric phase.  This is
readily seen in neutron inelastic scattering measurements made at
several temperatures above the Curie temperature.  In the top panel of
Fig.~1 we show the dispersion of the lowest-energy TO branch in
PbTiO$_3$ where at 20~K above $T_c$ the zone center ($\zeta = 0$)
energy has fallen to 3~meV \cite{Shirane}.

In relaxor compounds, however, there is a built-in disorder that
produces a diffuse phase transition in which the dielectric
permittivity $\epsilon$ exhibits a broad maximum as a function of
temperature at $T_{max}$.  In the case of PMN and PZN, both of which
have the simple $ABO_3$ perovskite structure, the disorder results
from the $B$-site being occupied by ions of differing valence (either
Mg$^{2+}$ or Zn$^{2+}$, and Nb$^{5+}$).  This breaks the translational
symmetry of the crystal.  Despite years of intensive research, the
physics of the observed diffuse phase transition is still not well
understood \cite{Westphal,Colla,Blinc}.  Moreover, it is interesting
to note that no definitive evidence for a soft mode has been found in
these systems.  The bottom panel of Fig.~1 shows neutron scattering
data taken by Naberezhnov {\it et al.} on PMN \cite{Naberezhnov}
exactly analogous to that shown in the top panel for PbTiO$_3$, except
that the temperature is $\sim 570$~K higher than $T_{max}$.

A seminal model for the disorder inherent to relaxors was first
proposed by Burns and Dacol in 1983 \cite{Burns}.  Using measurements
of the optic index of refraction on both ceramic samples of
(Pb$_{1-3x/2}$La$_x$)(Zr$_y$Ti$_{1-y}$)O$_3$ (PLZT) and single
crystals of PMN and PZN \cite{Burns}, they demonstrated that a
randomly-oriented local polarization $P_d$ develops at a well-defined
temperature $T_d$, frequently referred to as the Burns temperature,
several hundred degrees above the apparent transition temperature
$T_{max}$.  Subsequent studies have provided additional evidence of
the existence of $T_d$ \cite{Mathan,Bokov,Zhao}.  The spatial extent
of these locally polarized regions was conjectured to be $\sim$
several unit cells, and has given rise to the term ``polar
micro-regions,'' or PMR \cite{Tsurumi}.  For PZN-8\%PT, the formation
of PMR occurs at $T_d \sim$ 700~K, well above the cubic-to-tetragonal
phase transition at $T_c \sim$ 450~K.  We find striking anomalies in
the TO phonon branch (the same branch that goes soft at the zone
center at $T_c$ in PbTiO$_3$) that we speculate are directly caused by
these PMR.

%
\vspace{0.15in} 
\noindent 
\parbox[b]{3.4in}{ \psfig{file=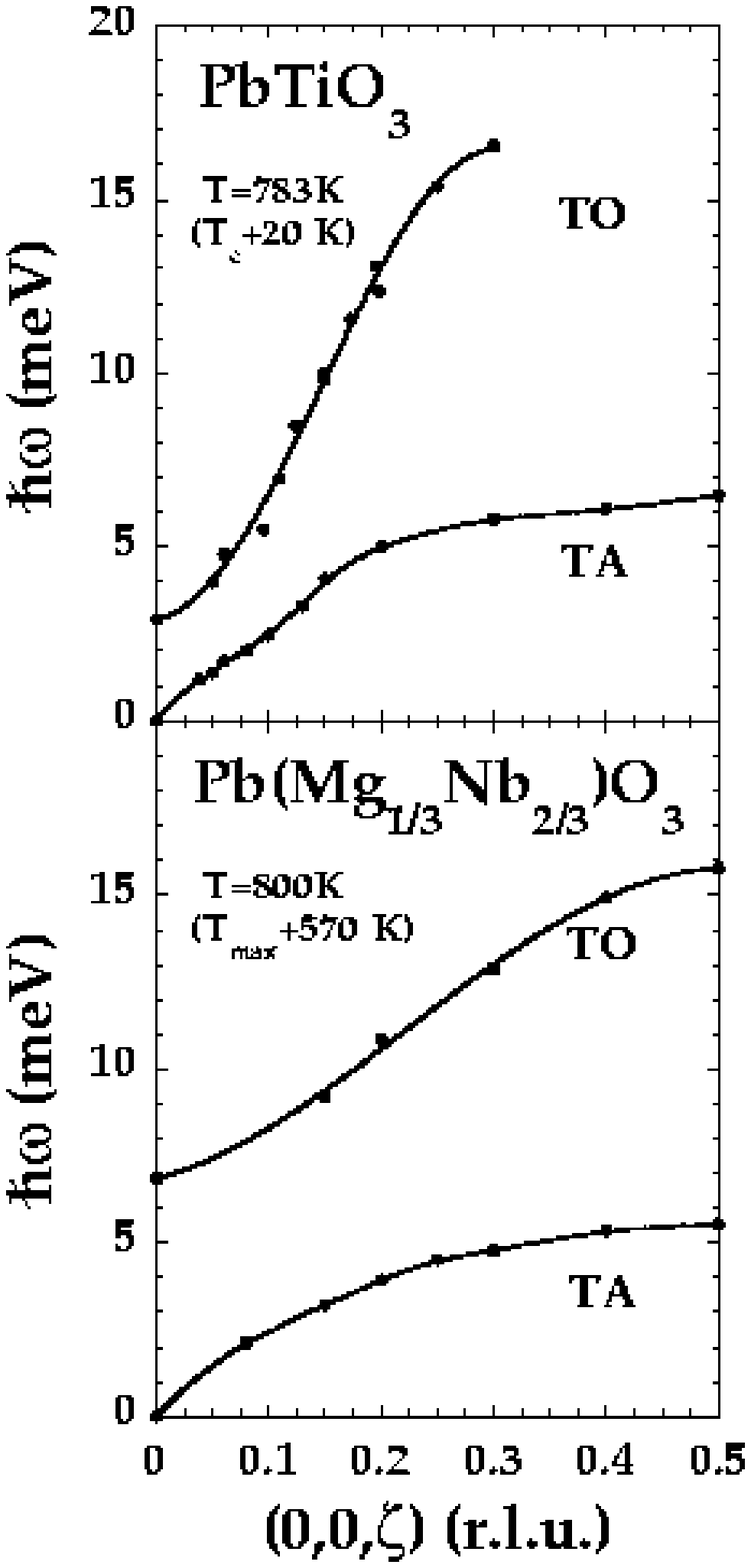,width=2.85in} {Fig.~1. \small
    Top - Dispersion of the lowest energy TO mode and the TA mode in
    PbTiO$_3$, measured just above $T_c$ (from Ref.\ \cite{Shirane}).
    Bottom - Dispersion curves of the equivalent modes in PMN measured
    far above $T_g$ (from Ref.\ \cite{Naberezhnov}). }}
\vspace{0.05in}
%

All of the neutron scattering experiments were performed on the BT2
and BT9 triple-axis spectrometers located at the NIST Center for
Neutron Research. The (002) reflection of highly-oriented pyrolytic
graphite (HOPG) was used to monochromate and analyze the incident and
scattered neutron beams.  An HOPG transmission filter was used to
eliminate higher-order neutron wavelengths.  The majority of our data
were taken holding the final neutron energy $E_f$ fixed at 14.7~meV
($\lambda_f = 2.36$~\AA) while varying the incident neutron energy
$E_i$, and using horizontal beam collimations
60$'$-40$'$-S-40$'$-40$'$.  The single crystal of PZN-8\% PT used in
this study weighs 2.8 grams and was grown using the high-temperature
flux technique described elsewhere \cite{Park2}.  The crystal was
mounted onto an aluminum holder and oriented with the either the cubic
[$\bar{1}$10] or [001] axis vertical.  It was then placed inside a
vacuum furnace capable of reaching temperatures up to 670~K.


%
\vspace{0.15in} 
\noindent 
\parbox[b]{3.4in}{ \psfig{file=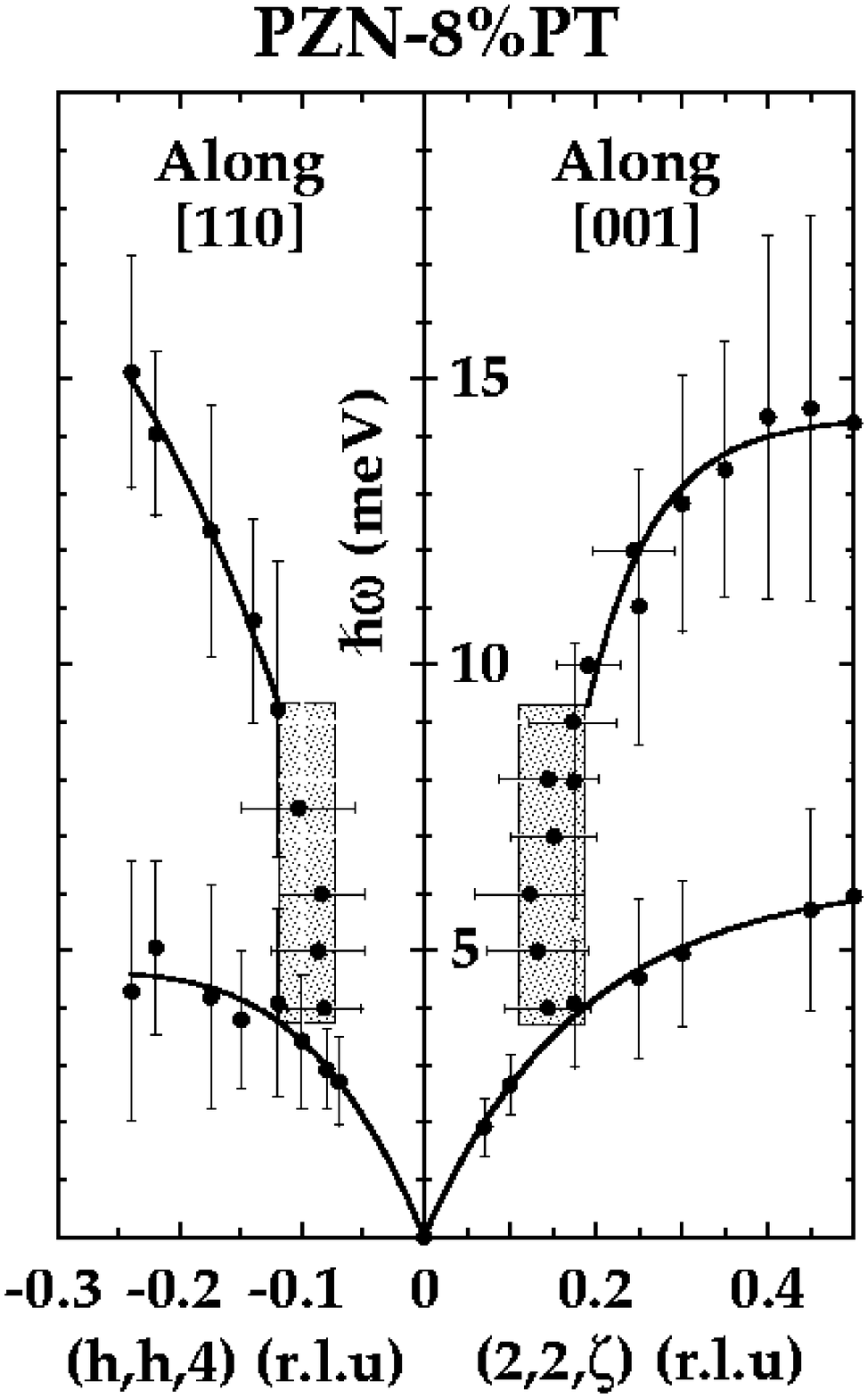,width=2.95in} {Fig.~2. \small
    Solid dots represent positions of peak scattered neutron intensity
    taken from constant-$\vec{Q}$ and constant-E scans at 500~K along
    both [110] and [001] symmetry directions.  Vertical (horizontal)
    bars represent phonon FWHM linewidths in $\hbar \omega$ ($q$).
    Solid lines are guides to the eye indicating the TA and TO phonon
    dispersions. }}
\vspace{0.05in}
%

Two types of scans were used to collect data.  Constant energy scans
were performed by keeping the energy transfer $\hbar \omega = \Delta E
= E_f - E_i$ fixed while varying the momentum transfer $\vec{Q}$.
Constant-$\vec{Q}$ scans were performed by holding the momentum
transfer $\vec{Q} = \vec{k_f} - \vec{k_i}$ ($k = 2\pi/\lambda$) fixed
while varying the energy transfer $\Delta E$.  Using these scans, the
dispersions of both the transverse acoustic (TA) and the lowest-energy
transverse optic (TO) phonon modes were mapped out at a temperature of
500~K (still in the cubic phase, but well below the Burns temperature
of $\sim$ 700~K).  In Fig.~2 we plot where the peak in the scattered
neutron intensity occurs as a function of $\hbar \omega$ and
$\vec{q}$, where $\vec{q} = \vec{Q} - \vec{G}$ is the momentum
transfer measured relative to the $\vec{G} = (2,2,0)$ and $(4,0,0)$
Bragg reflections along the symmetry directions [001] and [110],
respectively.  The horizontal scales of the left and right halves of
the figure have been adjusted so that each corresponds to the same $q$
(\AA$^{-1}$) per unit length.  The sizes of the vertical and
horizontal bars represent the phonon FWHM (full width at half maximum)
linewidths in $\hbar \omega$ (meV) and $q$ (\AA$^{-1}$), respectively,
and were derived from Gaussian least-squares fits to the
constant-$\vec{Q}$ and constant-$E$ scans.  The lowest energy data
points trace out the TA phonon branch along [110] and [001].  Solid
lines have been drawn through these points as a guide to the eye, and
are nearly identical to that shown for PMN in Fig.~1.

By far the most striking feature in Fig.~2 is the unexpected collapse
of the TO mode near the zone center where the polar optic branch
appears to drop precipitously, like a waterfall, into the acoustic
branch.  This anomalous behavior, shown by the shaded regions in
Fig.~2, stands in stark contrast to that of PMN at high temperature
where the same phonon branch intercepts the $\hbar \omega$-axis at a
finite energy (see bottom panel of Fig.~1).  The strange drop in the
TO phonon energy occurs for $q \sim 0.13$ r.l.u. measured along [001],
and for $q \sim 0.08$ r.l.u. measured along [110] (1 r.l.u. = $2\pi/a$
= 1.54~\AA$^{-1}$).  It is quite intriguing to note that these
$q$-values are both approximately equal to 0.2 \AA$^{-1}$.

To clarify the nature of this unusual observation, we show an extended
constant-$E$ scan taken at $\Delta E = 6$~meV in Fig.~3 along with a
constant-$\vec{Q}$ scan in the insert.  Both scans were taken at the
same temperature of 500~K, near the (2,2,0) Bragg peak, and along the
[001] direction.  The small horizontal bar shown under the left peak
of the constant-$E$ scan represents the instrumental FWHM
$q$-resolution, and is clearly far smaller than the instrinsic peak
linewidth.  We see immediately that the constant-$\vec{Q}$ scan shows
no evidence of any well-defined phonon peak, most likely because the
phonons near the zone center are overdamped.  However, the
constant-$E$ scan indicates the presence of a ridge of scattering
intensity at $\zeta = q = 0.13$~r.l.u., or about 0.2~\AA$^{-1}$, that
sits atop the scattering associated with the overdamped phonons.  Thus
the sharp drop in TO branch that appears to take place in Fig.~2 does
not correspond to a real dispersion curve as such.  Rather, it simply
indicates a region of ($\hbar \omega, q$)-space in which the phonon
scattering cross section is enhanced.  The origin of this enhancement
is unknown, however we speculate that it is a direct result of the PMR
described by Burns and Dacol \cite{Burns}.  If the length scale
associated with this enhancement is of order $2\pi/q$, this
corresponds to $\sim 31$~\AA, or about 7 to 8 unit cells, consistent
with Burns and Dacol's conjecture.

%
\vspace{0.15in} 
\noindent 
\parbox[b]{3.4in}{ \psfig{file=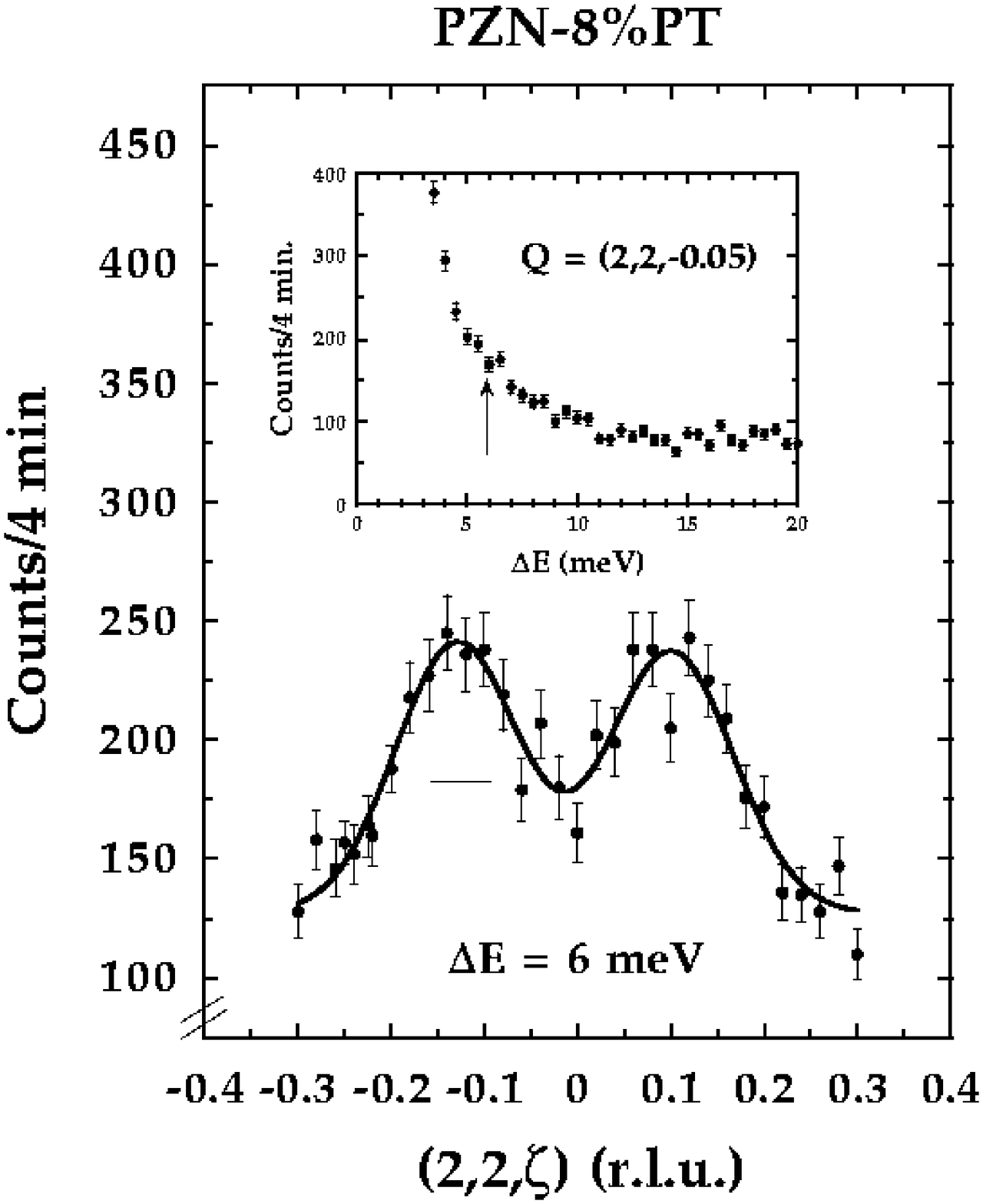,width=3.15in} {Fig.~3. \small
    Single constant-E scan measured along [001] at 6~meV at 500~K near
    the (2,2,0) Bragg peak.  Solid line is a fit to a double Gaussian
    function of $\zeta$.  The inset shows no peak in the scattered
    intensity measured along the energy axis.  The arrow indicates the
    position of the constant-E scan. }}
\vspace{0.05in}
%

Limited data were also taken as a function of temperature to determine
the effect on this anomalous ridge of scattering.  In Fig.~4 we show
two constant-$E$ scans, both measured at an energy transfer $\Delta E
= 5$~meV along the [010] direction, with one taken at 450~K, and the
other at 600~K.  The solid and dashed lines are fits to simple
Gaussian functions of $q$.  As is clearly seen, the ridge of
scattering shifts to smaller $q$, i.e. {\it towards} the zone center,
with increasing temperature.  These data strongly suggest a picture,
shown schematically in the inset to Fig.~4, in which the ridge of
scattering evolves into the expected classic TO phonon branch behavior
at higher temperature.  A single data point, obtained briefly at 670~K
to avoid damaging the crystal, is plotted in the inset to Fig.~4, and
tentatively corroborates this picture.

We have discovered an anomalous enhancement of the polar TO phonon
scattering cross section that occurs at a special value of $q =
0.2$~\AA$^{-1}$, independent of whether we measure along the [001] or
[110] direction.  We believe this to be direct microscopic evidence of
the PMR proposed by Burns and Dacol \cite{Burns}.  The presence of
such small polarized regions of the crystal above $T_c$ should
effectively prevent the propagation of long-wavelength ($q \rightarrow
0$) soft mode phonons.  A similar conclusion was reached by Tsurumi
{\it et al.} based on dielectric measurements of PMN \cite{Tsurumi}.
The observation that the phonon scattering cross section is enhanced
0.2~\AA$^{-1}$ from the zone center gives a measure of the size of the
PMR consistent with the estimates of Burns and Dacol.  If true, then
this unusual behavior should be observed in other related relaxor
systems.  Indeed, tentative evidence for this has already been
observed at room temperature in neutron scattering measurements on PMN
\cite{Gehring}.  This enhancement should also be reflected in x-ray
diffuse scattering intensities (\cite{Vakhrushev,You}), although it
may be masked by the superposition of strong acoustic modes.

%
\vspace{0.15in} 
\noindent 
\parbox[b]{3.4in}{ \psfig{file=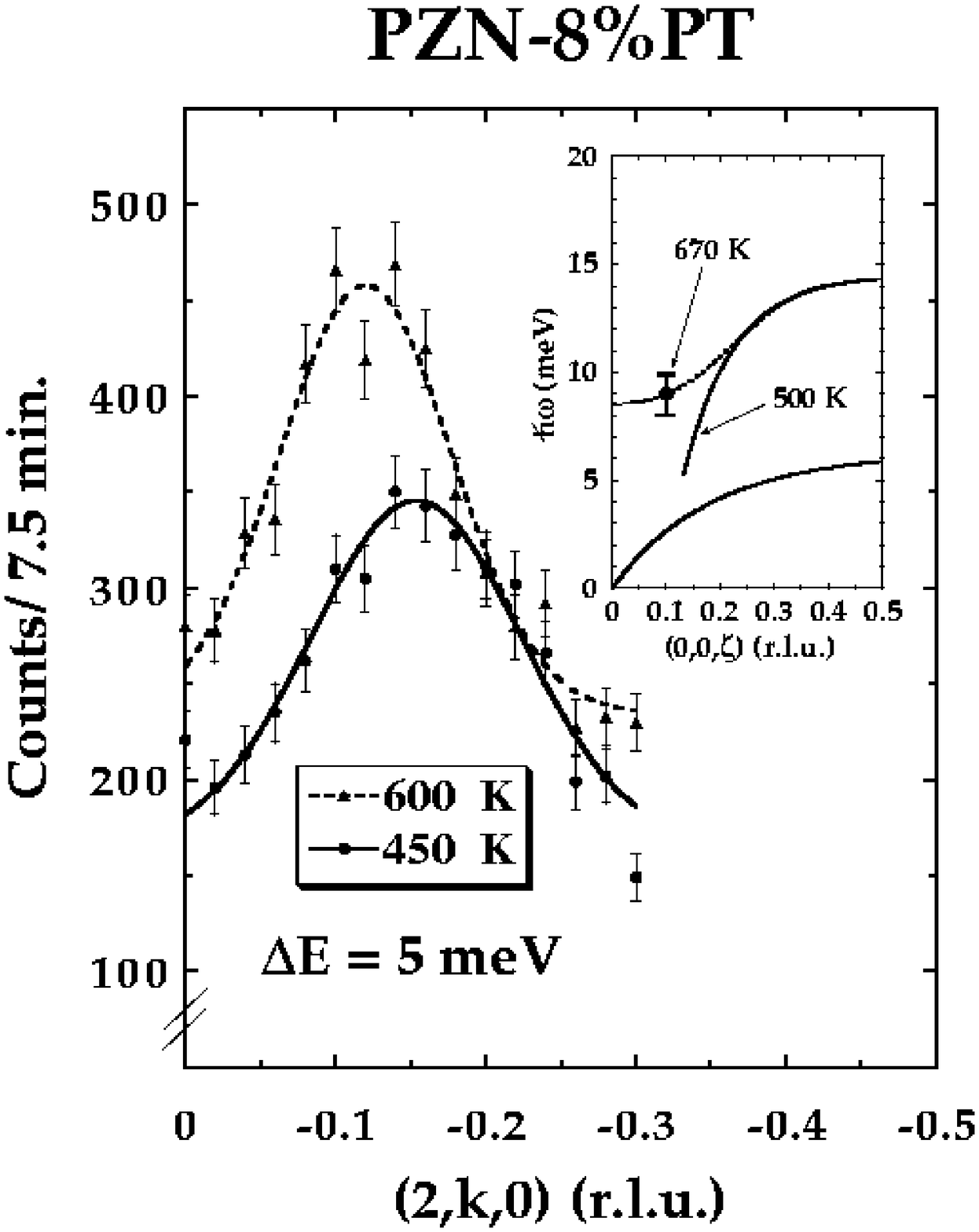,width=2.95in} {Fig.~4. \small
    Two constant-E scans measured along [010] at 5~meV at different
    temperatures.  The peak shifts towards the zone center with
    increasing temperature.  The inset suggests schematically how the
    TO branch dispersion recovers at higher temperatures. }}
\vspace{0.05in}
%

Our picture is not yet complete.  Whereas Fig.~3 demonstrates that
these anomalies appear as ridges on top of a broad overdamped cross
section, the complete nature of this cross section can only be
revealed by an extensive contour map of the Brillouin zone, for which
we lack sufficient data.  Another important aspect which requires
further study is exactly how the ``waterfall" evolves, at much higher
temperatures, into the standard optic mode dispersion as shown in
Fig.~1 for PMN.  We have not yet carried out this experiment because
of the concern of possible crystal deterioration at these high
temperatures under vacuum \cite{Park2}.  We intend to do so only after
all other key experiments have been completed \cite{X}.

Our current picture suggests that the TO phonon dispersion should
change if one alters the state of the PMR.  It is known that a macro
ferroelectric phase can be created in these relaxor crystals by
cooling the crystal in a field, or by application, at room
temperature, of a sufficiently strong field.  We are now planning
neutron inelastic measurements on such a crystal, as well as on PZN.

We thank S.\ Vakhrushev, S.\ Wakimoto, as well as D.\ E.\ Cox, L.\ E.\ 
Cross, R.\ Guo, B.\ Noheda, N.\ Takesue, and G.\ Yong for stimulating
discussions. Financial support by the U.\ S.\ Dept.\ of Energy under
contract No.\ DE-AC02-98CH10886, by ONR under project MURI
(N00014-96-1-1173), and under resource for piezoelectric single
crystals (N00014-98-1-0527) is acknowledged.  We acknowledge the
support of NIST, U.\ S.\ Dept.\ of Commerce, in providing the neutron
facilities used in this work.

\end{document}